\documentclass[12pt]{report}
\usepackage{epsf,amsmath}
\usepackage{indentfirst}
\usepackage{amssymb}
\usepackage{array,tabularx}
\begin{document}

\title{\large \bf The Mass of the Compact Object
in the Low-Mass X-ray Binary 2S 0921-630}

\author{\large\bf M. K. Abubekerov$^1$, E. A. Antokhina $^1$, A. M. Cherepashchuk$^1$,\\
\large\bf V. V. Shimanskii$^2$ \\
\normalsize\it $^1$ Sternberg Astronomical Institute, Russia \\
\normalsize\it $^2$ Kazan State University, Russia \\
}

\date{\begin{minipage}{15.5cm} \small
We interpret the observed radial-velocity curve of the optical
star in the low-mass X-ray binary 2S 0921-630 using a Roche model,
taking into account the X-ray heating of the optical star and
screening of X-rays coming from the relativistic object by the
accretion disk. Consequences of possible anisotropy of the X-ray
radiation are considered.We obtain relations between the masses of
the optical and compact (X-ray) components, $m_v$ and $m_x$, for
orbital inclinations $i=60^{\circ}$, $75^{\circ}$ è $90^{\circ}$.
Including X-ray heating enabled us to reduce the compact object's
mass by $\sim0.5-1M_{\odot}$, compared to the case with no
heating. Based on the K0III spectral type of the optical component
(with a probable mass of $m_v\simeq2.9M_{\odot}$), we concluded
that $m_x\simeq 2.45-2.55M_{\odot}$ (for
$i=75^{\circ}-90^{\circ}$). If the K0III star has lost a
substantial part of its mass as a result of mass exchange, as in
the V404 Cyg and GRS 1905+105 systems, and its mass is
$m_v\simeq0.65-0.75M_{\odot}$, the compact object's mass is close
to the standard mass of a neutron star, $m_x\simeq1.4M_{\odot}$
(for $i=75^{\circ}-90^{\circ}$). Thus, it is probable that the
X-ray source in the 2S 0921-630 binary is an accreting neutron
star.
\end{minipage}
} \maketitle \rm

\section*{\normalsize INTRODUCTION}

\quad Currently, mass estimates are available for $\sim 30$
neutron stars. The most accurate of these are the masses of radio
pulsars in Hulse-Taylor binaries \cite{Thorsett1999} and of the
radio pulsar J0737-3039 \cite{Lyne2004}. According to these
estimates, the masses of neutron stars are confined to a fairly
narrow range, $1.25-1.44M_{\odot}$. Nevertheless, theoretical
arguments suggesting the possible existence of massive neutron
stars with $m_{NS}\simeq 2-3M_{\odot}$ have accumulated.

First, theoretical computations indicate that the expected mass of
a neutron star formed during the core collapse of a massive star
can lie in the range $1.0-1.8M_{\odot}$
\cite{Woosley2000}-\cite{Fryer2001}.

Second, there exist quite a few hard equations of state for
neutron matter for which the Oppenheimer-Volkoff mass exceeds
$1.8M_{\odot}$ \cite{Haensal2003}. We especially note in this
context recently published papers on so-called "Skyrmion stars".
In 1999, Ouyed and Butler \cite{Ouyed1999} considered an equation
of state based on the model of Skyrme \cite{Skyrme1962}. A
characteristic feature of neutron-star models based on Skyrme's
equation is their high upper mass limit: $2.95M_{\odot}$ for
nonrotating and $3.45M_{\odot}$ for rotating stars
\cite{Ouyed2001,Ouyed2004}.

Third, evolutionary computations of neutron-star masses in binary
systems using the "Scenario Machine"\; \cite{Lipunov1996} have
demonstrated that there could exist binary-system evolutionary
channels, for which a neutron star is able to increase its mass
via accretion by more than $\sim 1M_{\odot}$
\cite{Bogomazov2005,Popov2005}.

Because of these arguments, special interest has been attracted by
the compact object in the low-mass X-ray binary 2S 0921-630
\cite{Li1978}. This binary X-ray system consists of a compact
object and a low-mass optical star (V395 Car)
\cite{Branduardi1983} of spectral type K0III \cite{Shahbaz1999}.
The binary's orbital period is $P_{orb}=9^{d}.006\pm0.007$
\cite{Jonker2005}. The nature of the compact object remains
unknown: the binary does not exhibit phenomena characteristic of
an X-ray pulsar or type 1 X-ray burster. Partial optical and X-ray
eclipses are observed for the system, testifying to a high orbital
inclination: $i\simeq70^{\circ}-90^{\circ}$
\cite{Shahbaz1999,Frank1987}.

According to Shahbaz et al.\cite{Shahbaz1999}, if the orbital
inclination is $i=70^{\circ}-90^{\circ}$, the compact object's
mass is $2.0-4.3M_{\odot}$. The component-mass ratio, derived from
the rotational line broadening, is $q=m_x/m_v=1.12\pm0.18$
\cite{Shahbaz1999}.

Jonker et al. \cite{Jonker2005} present the results of fitting an
accurate radial-velocity curve. Assuming
$i=60^{\circ}-90^{\circ}$, the mass of the compact object was
found to be $1.90\pm0.25<m_x<2.9\pm0.4M_{\odot}$
\cite{Jonker2005}. Given strong X-ray heating, these authors
applied a so-called K-correction \cite{Wade1988} to the
semiamplitude of the radial-velocity curve in order to estimate
the component-mass ratio, $q=m_x/m_v=0.75\pm0.37$
\cite{Jonker2005}.

The strength of the heating of the optical star's atmosphere by
X-rays from the compact object $k_x=L_x/L_v\simeq10$ complicates
the fitting of the radial-velocity curve for the 2S 0921-630
system. The model of the binary system used to analyze the
observed radial-velocity curve must take into account a variety of
physical phenomena related to the compact object's high X-ray
luminosity.

Keeping in mind the importance of reliable mass estimates for the
compact object in the 2S 0921-630 binary, we analyzed the
high-precision data of \cite{Jonker2005} using the Roche model of
\cite{Antokhina1994,Antokhina1996}, taking into account the X-ray
heating of the optical star, the screening of X-rays by the
accretion disk, and the possible anisotropy of the X-ray radiation
from the accretion disk associated with a black hole.

\section*{\normalsize OBSERVATIONAL MATERIAL}

We used the spectroscopic data of \cite{Jonker2005} for our
analysis. These data were acquired between December 2003 and March
2004 with the VLT (Very Large Telescope). A total of 44 spectra
with exposure times of 1300 s were obtained. The width of the
spectrograph's slit was $0^{''}.4$, making it possible to take
spectra with a high resolution ($0.75$ \AA\, per pixel). The
calibration uncertainty for the wavelength scale was 0.03 \AA.

Spectroscopic data were obtained using two gratings, 1200R+93 and
1028z+29, which covered 5920-6520 \AA\, and 8360-8900 \AA\,
respectively. The analysis of the spectrograms of the optical star
in the 2S 0921-630 system performed in \cite{Jonker2005} confirmed
its spectral type to be K0III. Radial velocities were determined
by cross-correlating the spectra taken with the 1200R+93
diffraction grating with the spectra of a standard star. Stars
with spectral types between G5 and K7, whose spectra were obtained
with the Keck telescope with the same resolution, were used as
standard stars. We adopted the

\begin{figure*}
\vspace{0cm} \epsfxsize=0.99\textwidth
\epsfbox{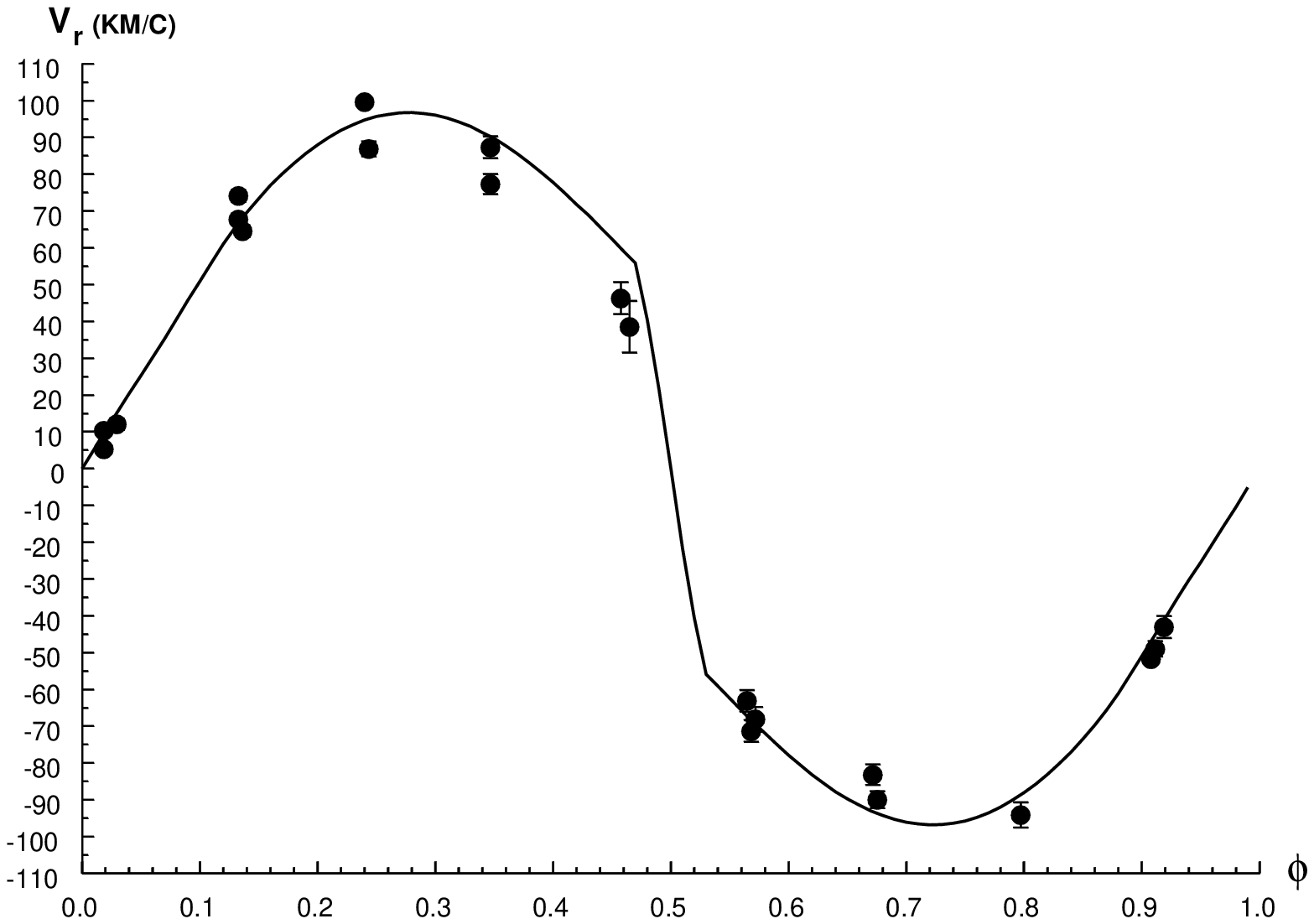} \caption{Observed
radial-velocity curve of the optical component of the X-ray binary
2S 0921-630 from \cite{Jonker2005} and the theoretical
radial-velocity curve in the Roche model, for the compact-object
mass $m_x=2.35M_{\odot}$, optical star mass $m_v=2.4M_{\odot}$,
X-ray heating coefficient for the optical star $k_x=10$, and
orbital inclination $i=75^{\circ}$ (the remaining binary
parameters are collected in Table \ref{param Roche})). The
theoretical radial-velocity curve corresponds to the
minimum-residual fit ($\chi_{min}^2=149.8$).} \label{Vobs}
\end{figure*}

We adopted the middle of the eclipse of the X-ray component by the
optical star to be zero phase when plotting the radial-velocity
curve. Jonker et al. \cite{Jonker2005} used their accurate
radial-velocity curve to refine the binary's orbital period, which
was found to be $P_{orb}=9^{d}.006\pm0.007$ relative to the zero
ephemeris $JD_o=2453000.49$. The system's spatial velocity is
$\gamma=44.4\pm2.4$ km/s. The semiamplitude of the observed
radial-velocity curve, obtained using the K1V radial-velocity
standard star HD 124106, is $K_v=99.1\pm3.1$ km/s. The observed
radial-velocity curve relative to the $\gamma$-velocity is shown
in Fig.\ref{Vobs}.

\section*{\normalsize ANALYSIS OF THE RADIAL-VELOCITY CURVES}

\quad The optical star in the 2S 0921-630 close binary fills its
Roche lobe. The star's surface is tidally and rotationally
distorted, and the side facing the relativistic component is
heated by X-rays from this component. The effects of the
component's interaction must be taken into account when analyzing
the optical component's radial-velocity curve. We fit the observed
radial-velocity curve using the Roche model, taking into account
the X-ray heating of the optical star. The algorithm we used is
described in detail by Antokhina et al. \cite{Antokhina1994} --
\cite{Antokhina2005}. We briefly summarize here the basic features
of this method.

\begin{table}[h!]
\caption{Numerical parameters used to model the optical
component's radial-velocity curve in the Roche model.}\label{param
Roche} \vspace{3.0mm} \centering
\def\arraystretch{1.0}
\newcolumntype{C}{>{\centering\arraybackslash}m}
\begin{tabular}{|l|l|p{110mm}|}
\hline
$P$(day)          & 9.006            & Orbital period \\
$m_v$($M_{\odot}$)& $var^{*}$        & Optical star's mass \\
$e$               & 0.0              & Eccentricity \\
$i(^\circ)$       & $60$, $75$, $90$ & Orbital inclination \\
$\mu$             & 1.0              & Roche lobe filling factor for the optical component  \\
$f$               & 1.0              & Rotational asynchronism factor for the optical component \\
$T_{\text{eff}}$(K)& 4700            & Optical component's effective temperature \\
$\beta$           & 0.08             & Gravity-darkening coefficient \\
$k_x$             & $var^{*}$        & Ratio of the relativistic
                                       component's X-ray luminosity to the optical component's bolometric
                                       luminosity, $L_x/L_{v}$ \\
$A  $             & 1.0              & Coefficient for reprocessing of the X-ray radiation \\
$u  $             & $0.3$            & Limb-darkening coefficient \\
$\alpha_p$        & $1.2$            & Photon index of the X-ray spectrum \\
\hline \multicolumn{3}{c}{}\\
\multicolumn{3}{c}{\begin{minipage}{15cm}\small $^{*}$ - These
parameters of the X-ray binary were varied in our model fitting.
\end{minipage}}\\
\end{tabular}
\end{table}

The binary consists of an optical star treated using the Roche
model and a point X-ray source. The optical star's surface is
subdivided into several thousand (in our case, $\sim2600$) area
elements, each of which we compute the emergent local flux
assuming LTE, including the effect of the incident X-ray flux.
Each area element is described by local temperature $T_{loc}$,
local gravity $g_{loc}$, and a parameter $k_{x}^{loc}$, equal to
the ratio of the incident X-ray flux to the emergent radiation
flux, without taking into account external irradiation of the
atmosphere. For these parameters at a given point on the surface,
we solved the radiative-transport equations in a spectral line in
the presence of the incident X-ray radiation, in order to compute
a model atmosphere and obtain the intensity of the emergent
radiation in the line and continuum. Our model synthesis of the
radial-velocity curve was performed for the CaI 6439.075 \AA\,
absorption line (hereafter, the CaI 6439 \AA\, line). For a given
phase of the orbital period, we summed the contributions of the
areas to the total flux, allowing for Doppler effects and the
visibility of the areas for the observer. In this way, we computed
the integrated continuum radiation flux from the star towards the
observer, as well as the rotation-broadened profile of the
spectral line, which was used to derive the star's radial
velocity. The radial velocity of the star at a given orbital phase
was determined from the mean wavelengths at residual intensity
levels for the integrated absorption-line profile of one-third,
one-half, and two-thirds.

We carried out our analysis both including and excluding the
effect of the instrumental profile on the model integrated profile
of the CaI 6439 \AA\, line. The full width at half-maximum of the
instrumental profile was taken to be FWHM=0.5\,\AA. The two sets
of results were virtually identical, and we accordingly present
here only those including the effect of the instrumental profile
on the integrated profile of the CaI 6439 \AA\, line.

We noted above that considerable X-ray heating of the optical
star's atmosphere was observed in the binary system. Thus, the
X-ray spectrum must be taken into account as accurately as
possible, using the known X-ray spectrum \cite{Kallman2003}. Based
on observations obtained with the XMM and Chandra observatories
\cite{Kallman2003}, we adopted a photon index for the power-law
spectrum $\alpha_p=1.2$ in the range 0.1-12 keV.

The compact object's intrinsic X-ray luminosity indicated by the
XMM and Chandra data is $L_x\sim 10^{36}$ erg/s
\cite{Kallman2003}. Assuming that the mean effective temperature
of the optical star filling its Roche lobe is $T_{eff}=4700$ K,
its luminosity is $L_{v}\simeq2\cdot10^{35}$ erg/s. Given the
uncertainties in the estimates of the distance to the system and
of $L_x$, we can assume $k_x=L_x/L_v\simeq10$. We fit the observed
radial-velocity curve using the atmospheric heating coefficients
$k_x=0$ and $k_x=10$.

We fit the observed radial-velocity curve in three different
models: with screening of the X-rays by the accretion disk both
taken and not taken into account, and by allowing for the
anisotropy of the X-rays emitted by the accretion disk of a
rotating black hole. The numerical parameters of the Roche model
for the 2S 0921-630 binary are presented in Table \ref{param
Roche}.

\section*{\normalsize\it Model 1. Analysis
of the Observed Radial-Velocity Curve without Screening of the
X-ray Flux by the Accretion Disk}

Since the exact mass of the optical star is not known, we treated
the masses of both binary components as parameters to be
determined. Since the inverse problem is rather cumbersome in this
formulation, we carried out an exhaustive search for the
parameters, thereby repeatedly solving the direct problem. For
each mass of the optical component $m_v$ from the discrete set of
values $1.0$, $1.9$, $2.4$, $2.9M_{\odot}$, fixing i, we carried
out an exhaustive search for the compact object's mass $m_x$. We
tested the model's fit to the observations using the usual
$\chi^2$ statistical test, working with the significance levels
$\alpha=1\%$ and 5\% (cf. \cite{Cherepashchuk1993} for details).

\small
\begin{table}
\caption{Dependence of mx on mv for the Roche model including
isotropic X-ray heating of the optical component with $k_x=10$ and
$i=60^{\circ}$, $75^{\circ}$, $90^{\circ}$. (without screening of
the X-rays by the accretion disk).}\label{kx10_nondisk}
\vspace{3.0mm}\centering
\def\arraystretch{1.4}
\newcolumntype{C}{>{\centering\arraybackslash}m}
\begin{tabular}{|C{14mm}|C{19mm}|C{19mm}|C{19mm}|}
\hline
 $m_v(M_{\odot})$ & $i=60^{\circ}$ $m_x(M_{\odot})$ & $i=75^{\circ}$ $m_x(M_{\odot})$ & $i=90^{\circ}$ $m_x(M_{\odot})$ \\
\hline
 1.0             & 2.00 & 1.70  & 1.55  \\
 1.9             & 2.55 & 2.15  & 2.00  \\
 2.4             & 2.80 & 2.35  & 2.25  \\
 2.9             & 3.00 & 2.55  & 2.45  \\
\hline
\end{tabular}
\end{table}
\normalsize

To quantitatively estimate the effect of X-ray heating on our
estimates of $m_x$, we fit the observed radial-velocity curve with
X-ray heating both taken and not taken into account, and also
allow for possible anisotropy of X-ray radiation from inner parts
of the accretion disk around a black hole. When X-ray heating was
taken into account, we set $k_x=10$. We obtained fits for
inclinations $i=60^{\circ}$, $75^{\circ}$ è $90^{\circ}$. This
analysis yields relations between the masses of the optical and
compact objects, $m_v$ and $m_x$. The numerical results are
collected in Tables \ref{kx10_nondisk} and \ref{kx0_nondisk}, and
are plotted in Fig. \ref{mxfmv}.

Tables \ref{kx10_nondisk} and \ref{kx0_nondisk} present the values
of $m_x$ corresponding to the minimum $\chi^2$ residuals for the
difference between the theoretical and observed radial-velocity
curves. The values of $m_x$ in Tables \ref{kx10_nondisk} and
3\ref{kx0_nondisk} are not accompanied by uncertainties, because
both of the binary models were rejected at the $\alpha=1\%$ and
5\% significance levels. The quantiles for these significance
levels are $\Delta_{1\%}=38.93$ and $\Delta_{5\%}=32.67$. When
X-ray heating of the optical star is included ($k_x=10$), the
minimum residual is $\chi_{min}^2\simeq120-150$. When it is not
included ($k_x=0$), the minimum residual increases to
$\chi_{min}^2\simeq350-380$. The statistical inadequacy of the
models in describing the observational data is due to the
considerable scatter of the data points, with uncertainties in the
radial velocities for each data point being comparatively small
(see Fig. \ref{Vteor} below). As was noted in \cite{Jonker2005},
the considerable scatter of the data points in the radial-velocity
curve is probably due to variable X-ray heating of the star.

Let us compare the results of our fitting of the observed
radial-velocity curve obtained using the Roche model both with
(Table \ref{kx10_nondisk}) and without (Table \ref{kx0_nondisk})
X-ray heating. With X-ray heating ($k_x=10$), with
$i=60^{\circ}-90^{\circ}$ and $m_v=1.0-2.9M_{\odot}$, the mass of
the relativistic object is $m_x=1.55-3.0M_{\odot}$; without X-ray
heating ($k_x=0$), the compact object's mass is
$m_x=1.95-4.0M_{\odot}$. Thus, the fit of the data without X-ray
heating of the optical star's atmosphere leads to masses of the
compact object that are too high by $0.5-1.0M_{\odot}$

Let us consider in more detail why $m_x$ is systematically too
high in the model without X-ray heating. Figure \ref{CaIphase025}
displays the model integrated profiles of the CaI 6439\,\AA\, line
at orbital phase $\phi=0.25$ obtained for $m_v=2.4M_{\odot}$,
$m_x=2.35M_{\odot}$, and $i=75^{\circ}$, with X-ray heating both
taken (solid curve) and not taken (dashed curve) into account.

When X-ray heating is taken into account, an emission component
appears in the integrated profile of the absorption line (Fig.
\ref{CaIphase025}). Because of this emission component, the
"center of gravity"\, of the absorption line is displaced. Thus,
when the X-ray heating is included, the line's center of gravity
in Fig. \ref{CaIphase025} is at a wavelength of 6441.131\,\AA,
corresponding to a radial velocity of 95.44 km/s. If there is no
X-ray heating, the line's center of gravity is at 6440.818\,\AA,
corresponding to a radial velocity of 81.57 km/s. Thus, due to the
distortion of the absorption line profile by the emission
component, the observed amplitude of the optical component's
radial-velocity curve can be explained using a lower mass for the
compact object. For this reason, carefully taking into account the
X-ray heating in the 2S 0921-63 binary reduces the compact
object's mass from $m_x=1.95-4.0M_{\odot}$ to
$m_x=1.55-3.0M_{\odot}$.

\small
\begin{table}
\caption{ Dependence of mx on mv for the Roche model without
heating the optical component ($k_x=0$) for $i=60^{\circ}$,
$75^{\circ}$, $90^{\circ}$.}\label{kx0_nondisk}
\vspace{3.0mm}\centering
\def\arraystretch{1.4}
\newcolumntype{C}{>{\centering\arraybackslash}m}
\begin{tabular}{|C{14mm}|C{19mm}|C{19mm}|C{19mm}|}
\hline
 $m_v(M_{\odot})$ & $i=60^{\circ}$ $m_x(M_{\odot})$ & $i=75^{\circ}$ $m_x(M_{\odot})$ & $i=90^{\circ}$ $m_x(M_{\odot})$ \\
\hline
 1.0             & 2.55 & 2.10  & 1.95  \\
 1.9             & 3.30 & 2.75  & 2.60  \\
 2.4             & 3.65 & 3.05  & 2.85  \\
 2.9             & 4.00 & 3.35  & 3.15  \\
\hline
\end{tabular}
\end{table}
\normalsize

This is illustrated by model radial-velocity curves in
\ref{Vteor}. For X-ray heating with $k_x=10$, $m_v=2.4M_{\odot}$,
$i=75^{\circ}$, the minimum residual ($\chi_{min}^2=149.8$) is
achieved for a mass of the compact object $m_x=2.35M_{\odot}$
(solid curve in Fig.\ref{Vteor}). The semiamplitude of the optical
componentfs radialvelocity curve is $K_v=96.81$ km/s. For the
same binary parameters but without X-ray heating, the
radial-velocity curve has a semiamplitude of $K_v=81.27$ km/s
(dash-dotted curve in Fig.\ref{Vteor}). For the analysis without
X-ray heating and with $m_v=2.4M_{\odot}$ and $i=75^{\circ}$, the
minimum residual ($\chi_{min}^2=356.7$) is achieved for the higher
mass of the compact object $m_v=3.05M_{\odot}$ (dashed curve in
Fig.\ref{Vteor}).

Let us also consider the shape of the optical star's absorption
line in the presence of strong X-ray heating. It is obvious that
estimates of the mass of the compact object depend directly on the
uncertainty in the center of gravity of the absorption line in the
optical star's spectrum. When analyzing the shape of model
integrated profiles for the CaI 6439\,\AA\, line, we encountered
the problem of determining the line's center of gravity.

For example, in a binary with $m_v=2.4M_{\odot}$,
$m_v=2.35M_{\odot}$, and $i=75^{\circ}$, the center of gravity of
the line at phases 0.47-0.50 cannot be determined unambiguously.
Fig.\ref{CaIphase045050} presents integrated profiles of the CaI
6439\,\AA\, line for orbital phases 0.45-0.50. We can see that,
beginning with orbital phase 0.47, the line profile has two
minima, preventing unambiguous determination of the center of
gravity of the line, and hence of the radial velocity. For this
reason, a break in the radial-velocity curve is observed at phases
0.47 and 0.53 (Fig. \ref{Vteor}).

\small
\begin{table}
\caption{Dependence of $m_x$ on $m_v$ for the Roche model with
$i=75^{\circ}$, heating of the optical component with $k_x=10$,
and screening of the X-rays by the accretion
disk.}\label{kx10_disk} \vspace{3.0mm}\centering
\def\arraystretch{1.4}
\newcolumntype{C}{>{\centering\arraybackslash}m}
\begin{tabular}{|C{25mm}|C{25mm}|}
\hline
 $m_v(M_{\odot})$ & $m_x(M_{\odot})$ \\
\hline
 1.0             & 1.70 \\
 1.9             & 2.15 \\
 2.4             & 2.40 \\
 2.9             & 2.60 \\
\hline
\end{tabular}
\end{table}
\normalsize

Because of the ambiguity in the absorption-line profiles in X-ray
binary systems whose optical stars are heated, with
$k_x\gtrsim10$, it is difficult to derive a single satisfactory
radial-velocity curve. Consequently, the masses of the
relativistic objects must be estimated directly from the orbital
variations of the observed line profiles, rather than from
indirect data, such as radial-velocity curves. Modern 8-10m
telescopes can provide the needed high quality observational data
for this purpose.

\section*{\normalsize\it Model 2. Analysis
of the Observed Radial-Velocity Curve with Screening of the X-ray
Flux by the Accretion Disk}

In this section, we allow for screening of the X-ray flux by the
accretion disk, assuming the total screening angle of the
accretion disk to be $5^{\circ}$ \cite{Shakura1973} and the
accretion disk to be situated in the binary's orbital plane. Some
area elements on the optical star happen to be in the strip that
is shielded from the X-rays from the relativistic object. We
computed absorption-line intensities for these areas proceeding as
above, taking the local X-ray heating coefficient to be
$k_{x}^{loc}=0$. We analyzed the observed radial-velocity curve
for discretemasses of the optical star $m_v=1.0$, $1.9$, $2.4$,
$2.9M_{\odot}$, and for $i=75^{\circ}$. The results are presented
in Table \ref{kx10_disk}.

The results were quite close to (for masses $m_v=1.0$), and
$1.9M_{\odot}$, identical with those obtained in the Roche model
without screening the X-ray flux by the accretion disk (Table
\ref{kx10_nondisk}). For this reason, we did not continue our
analysis for orbital inclinations $i=60^{\circ}$ and $90^{\circ}$.
The shadow strip on the optical component turned out to be too
narrow to have any considerable influence on the results.

Similarly, we analyzed the observed radial-velocity curve allowing
for screening of the X-ray flux by the accretion disk for the case
$k_x=30$, with $i=75^{\circ}$. and $m_v=1.0$, $1.9$, $2.4$,
$2.9M_{\odot}$, as in the previous case. The results are presented
in Table \ref{kx30_disk}. The derived masses are tabulated without
error intervals, because all the binary models were rejected at
the 1\% and 5\% significance levels. The $m_x$ values given
correspond to the minimum $\chi^2$ residuals.

\small
\begin{table}
\caption{Dependence of $m_x$ on $m_v$ for the Roche model with
$k_x=30$ and $i=75^{\circ}$.}\label{kx30_disk}
\vspace{3.0mm}\centering
\def\arraystretch{1.4}
\newcolumntype{C}{>{\centering\arraybackslash}m}
\begin{tabular}{|C{20mm}|C{50mm}|C{50mm}|}
\hline
 $m_v(M_{\odot})$ & with screening of the X-rays $m_x(M_{\odot})$ & without screening of the X-rays $m_x(M_{\odot})$ \\
\hline
 1.0             & 1.65  & 1.60  \\
 1.9             & 2.08  & 2.05  \\
 2.4             & 2.30  & 2.28  \\
 2.9             & 2.50  & 2.45  \\
\hline
\end{tabular}
\end{table}
\normalsize

It follows from Table \ref{kx30_disk} that including screening of
the X-ray flux $L_x=30L_v$ by the accretion disk has little
influence on the results. The largest mass difference revealed for
the two models is $0.05M_{\odot}$ (Table \ref{kx30_disk}). The
effect of screening will be even smaller if the heating
coefficient is lower. Thus, the X-ray pulsar masses we derived
earlier in \cite{Abubekerov2004_1} that did not allow for
screening of the X-ray flux by the accretion disks can be
considered reliable.

\section*{\normalsize\it Analysis of the Observed Radial-Velocity Curve Allowing for Anisotropic
X-ray Radiation from the Accretion Disk}

To complete our study, we considered the hypothesis that the 2S
0921--630 binary contains a low-mass black hole. In this case, the
X-rays from the accretion disk should be anisotropic, reducing the
X-ray flux in the orbital plane. Note that the presence of strong
emission components for many absorption lines in the optical
component's spectrum \cite{Jonker2005} suggests that the X-ray
radiation is isotropic. The emission lines in the spectrum of the
optical star, which indicate considerable X-ray heating of the
optical component, also provide indirect evidence that the system
contains a neutron star.

We assumed the accretion disk to be optically thin and located in
the binary's orbital plane. The anisotropy of the X-rays from the
accretion disk was described with Eqs. (\ref{Lxray1}) and
(\ref{Lxray2}) from \cite{Bochkarev1988}:

\begin{equation}
\frac{dL}{d\Omega}=\frac{L_x\cdot F(\theta)}{4\pi}, \label{Lxray1}
\end{equation}

\begin{equation}
F(\theta)=\frac{6}{7}\cos\theta(1+2\cos\theta) , \label{Lxray2}
\end{equation}

\small
\begin{table}
\caption{Dependence of $m_x$ on $m_v$ for the Roche model with
$k_x=10$, anisotropic radiation from the X-ray source, and
$i=60^{\circ}$, $75^{\circ}$, $90^{\circ}$.}\label{kx10_ani}
\vspace{3.0mm}\centering
\def\arraystretch{1.4}
\newcolumntype{C}{>{\centering\arraybackslash}m}
\begin{tabular}{|C{14mm}|C{19mm}|C{19mm}|C{19mm}|}
\hline
 $m_v(M_{\odot})$ & $i=60^{\circ}$ $m_x(M_{\odot})$ & $i=75^{\circ}$ $m_x(M_{\odot})$ & $i=90^{\circ}$ $m_x(M_{\odot})$ \\
\hline
 1.0             & 2.20 & 1.80  & 1.70  \\
 1.9             & 2.75 & 2.30  & 2.15  \\
 2.4             & 3.00 & 2.50  & 2.35  \\
 2.9             & 3.25 & 2.75  & 2.60  \\
\hline
\end{tabular}
\end{table}
\normalsize

\noindent where $\theta$ is the angle between the normal to the
disk plane and the direction of an element of solid angle
$d\Omega$. The geometry of the binary system for the Roche model
with the parameters in Table \ref{param Roche} suggests that the
X-ray flux falls on the optical component at
$\theta\simeq70^{\circ}-90^{\circ}$. We can see from
(\ref{Lxray2}) that the X-ray flux is considerably attenuated for
such values of $\theta$. Figure \ref{CaIphase025_ani} presents the
model integrated profiles of the CaI 6439\,\AA\, line at orbital
phase 0.25. Note that the emission component of the integrated
line profile is weaker than in the case of an isotropic X-ray
flux.

The results of our fitting of the observed radialvelocity curve
for the case of anisotropic X-rays from the accretion disk are
presented in Table \ref{kx10_ani} and plotted in Fig.
\ref{mxfmv_ani}. We give the central values without uncertainties,
since the binary models were rejected at the $\alpha=1\%$ and
$5\%$ significance levels. For $m_v=1.0-2.9M_{\odot}$, the mass of
the black hole is $1.70-3.25M_{\odot}$. Recall that the compact
object's mass for the case of isotropic X-rays with $k_x=10$ was
$m_x=1.55-3.0M_{\odot}$. Thus, allowing for an anisotropic X-ray
flux from the accretion disk increases the compact object's mass
by $\sim0.2 M_{\odot}$ compared to the case of an isotropic X-ray
flux.

\section*{\normalsize DISCUSSION}

The results of our analysis of the high-precision radial-velocity
curve \cite{Jonker2005} (Fig. \ref{Vobs}) with orbital
inclinations $i=60^{\circ}-90^{\circ}$ and the optical component's
mass $m_v=1.0-2.9M_{\odot}$ are presented in Table
\ref{Result_mx}. We consider the mass of the compact object
obtained in Model 1 to be our main result.

As we noted above, the binary exhibits X-ray eclipses, testifying
to high orbital inclination: $i\simeq70^{\circ}-90^{\circ}$
\cite{Shahbaz1999}. Since there are no X-ray dips, we can assume
$i>80^{\circ}$ \cite{Frank1987}. For our mass estimates, we
adopted values $i=75^{\circ}-90^{\circ}$. In this case, assuming
$m_v=1.0-2.9M_{\odot}$, the mass of the compact object is
$m_x\simeq1.55-2.55M_{\odot}$ (Fig. \ref{mxfmv}). Taking into
account the optical component's spectral type, K0III
($m_v\simeq2.9M_{\odot}$ \cite{Straijis}), we find the compact
object's mass to be $m_x=2.45-2.55M_{\odot}$ (Fig. \ref{mxfmv}).

\small
\begin{table}
\caption{Mass of the compact object in the 2S 0921-630 X-ray
binary for various models, with $k_x=10$.}\label{Result_mx}
\vspace{3.0mm}\centering
\def\arraystretch{1.4}
\newcolumntype{C}{>{\centering\arraybackslash}m}
\begin{tabular}{|C{50mm}|C{50mm}|}
\hline
Ìîäåëü$^*$ & $m_x(M_{\odot})$ \\
\hline
 Ìîäåëü 1 & 1.55--3.0  \\
 Ìîäåëü 2 & 1.55--3.0  \\
 Ìîäåëü 3 & 1.70--3.25 \\
\hline \multicolumn{2}{c}{}\\
\multicolumn{2}{c}{\begin{minipage}{10cm}\small $^{*}$ -- Model 1:
isotropic X-ray source, without screening of the X-ray flux by the
accretion disk; model 2: isotropic X-ray source, with screening of
the X-ray flux by the accretion disk; model 3: anisotropic X-ray
source, without screening of the X-ray flux by the accretion disk.
\end{minipage}}\\
\end{tabular}
\end{table}
\normalsize

The actual mass of the optical star in an interacting binary,
$m_v$, can differ from the estimate obtained based on its spectral
type. For example, the mass estimates for the low-mass transient
X-ray binaries V404 Cyg (spectral type of the optical star K0IV)
and GRS 1915+105 (spectral type of the optical star KIII) derived
from rotational broadening of absorption lines are
$0.7\pm0.1M_{\odot}$ \cite{Casares1994} and $0.81\pm0.53M_{\odot}$
\cite{Greiner2004}, respectively. At the same time, mass estimates
based on the corresponding spectral types and luminosity classes
are $\sim 1.3M_{\odot}$ and $\sim 2.3-2.9M_{\odot}$
\cite{Straijis}, respectively. Thus, the estimates based on
spectral type are too high by almost a factor of $\sim 2$ compared
to the spectroscopic $m_v$ values. This is in agreement with
current theoretical ideas about stellar evolution in low-mass
close binaries. According to computations using the "Scenario
Machine" \cite{Lipunov1996}, an optical component with initial
mass $\sim 3M_{\odot}$ loses more than $\sim 1M_{\odot}$ during
the mass-exchange stage \cite{Bogomazov2005}.

The optical stars in the X-ray binaries V404 Cyg, GRS 1915+105,
and 2S 0921--630 have similar spectral types and luminosity
classes \cite{Shahbaz1999,Greiner2004}. Assuming that the K0III
star in the 2S 0921--630 system lost a substantial fraction of its
mass via exchange and that its mass, like the optical star in GRS
1905+105, is $m_v\simeq1M_{\odot}$, we find the mass of the
compact object in the 2S 0921--630 system to be
$m_x\simeq1.6-1.7M_{\odot}$ (for $i=75^{\circ}-90^{\circ}$). This
is close to the mean mass of a neutron star, making it likely that
we are dealing with an accreting neutron star in the 2S 0921--630
binary.

To extend our analysis for the 2S 0921-630 binary, we considered
the possibility that the compact object's mass is close to the
standard mass of a neutron star, $m_x=1.4M_{\odot}$. In this case,
the mass of the optical component is $m_v=0.65-0.75M_{\odot}$,
where we have assumed $i=75^{\circ}-90^{\circ}$ since partial
X-ray eclipses are observed for the system \cite{Shahbaz1999}.
Formally, from the stellar-evolution point of view, single stars
with such masses cannot become giants over the Hubble time. In
this case, the presence of a relativistic companion enables a star
with an initial mass $\gtrsim 0.8M_{\odot}$ to lose some its mass
during the semidetached phase \cite{Tutukov2003}.

The result of the evolution of a giant with mass
$\sim0.6-0.7M_{\odot}$ is a helium white dwarf with a mass
$\sim0.3-0.4M_{\odot}$ \cite{Tutukov2003}. Thus, it is quite
plausible that further evolution of the 2S 0921-630 X-ray binary
will result in a millisecond radio pulsar in a binary with a
helium white dwarf. This hypothesis is in good quantitative
agreement with the model computations of \cite{Tutukov2003} and
the empirical relation between the orbital periods and secondary
(white dwarf) masses for binaries with millisecond radio pulsars
(cf. [\cite{Tutukov2003}, Fig. 4] and [\cite{Lorimer2005}, Fig.
9]).

We note again that we rejected all models for the X-ray binary
system at the $\alpha=1\%$ and 5\% significance levels. Tables
\ref{kx10_nondisk} -- \ref{Result_mx} present the $m_x$ values
corresponding to the minimum $\chi^2$ residuals. We must bear in
mind that, in all cases, the models do not completely adequately
represent the observational data, so that our estimates of $m_x$
cannot be considered final.

Note the importance of the methodological results of our study.
Our computations for the Roche model demonstrated that the profile
of the absorption line is rather complex in the case of strong
X-ray heating, and suffers considerable variations in the course
of orbital motion (Figs. \ref{CaIphase025}, \ref{CaIphase045050}).
Thus, estimates of $q$ based on rotational broadening of
absorption-line profiles in low-mass X-ray binaries with strong
X-ray heating cannot be considered trustworthy. Our future plans
include a study of this problem.

\section*{\normalsize CONCLUSIONS}

\begin{figure*}[h!]
\vspace{0cm} \epsfxsize=0.99\textwidth \epsfbox{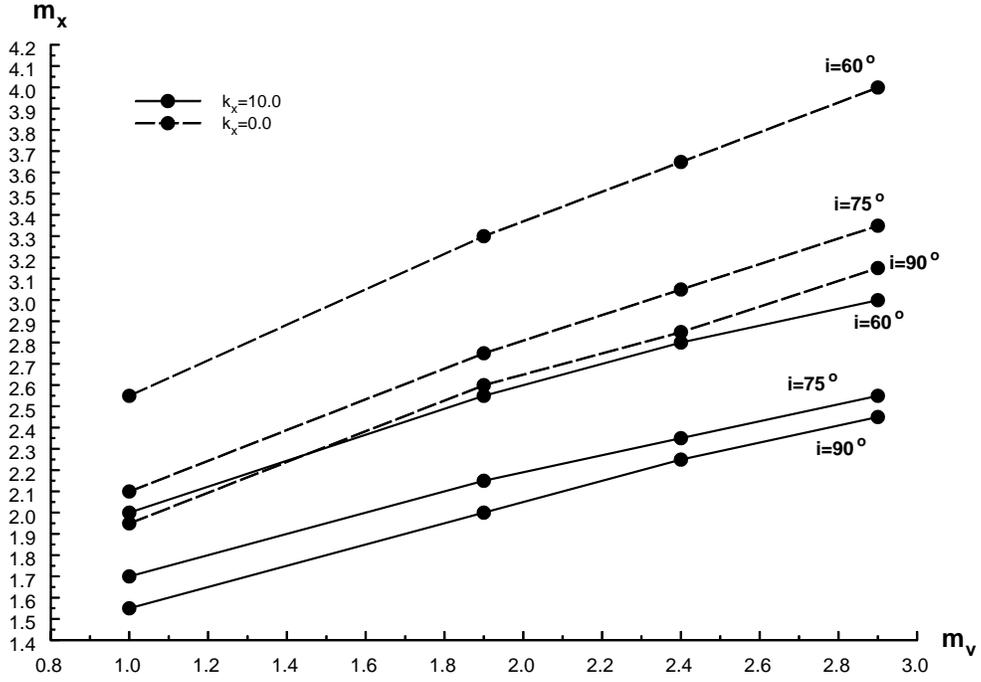}
\caption{\small Relations between the masses of the optical star
and the relativistic component in the 2S 0921-63 binary from our
fitting of the observed radial-velocity curve using the Roche
model, assuming $k_x=10$ (solid broken lines) and $k_x=0$ (dashed
broken lines) for X-ray heating. The X-ray heating is taken to be
isotropic.}\label{mxfmv}
\end{figure*}

\begin{figure*}
\vspace{0cm} \epsfxsize=0.99\textwidth
\epsfbox{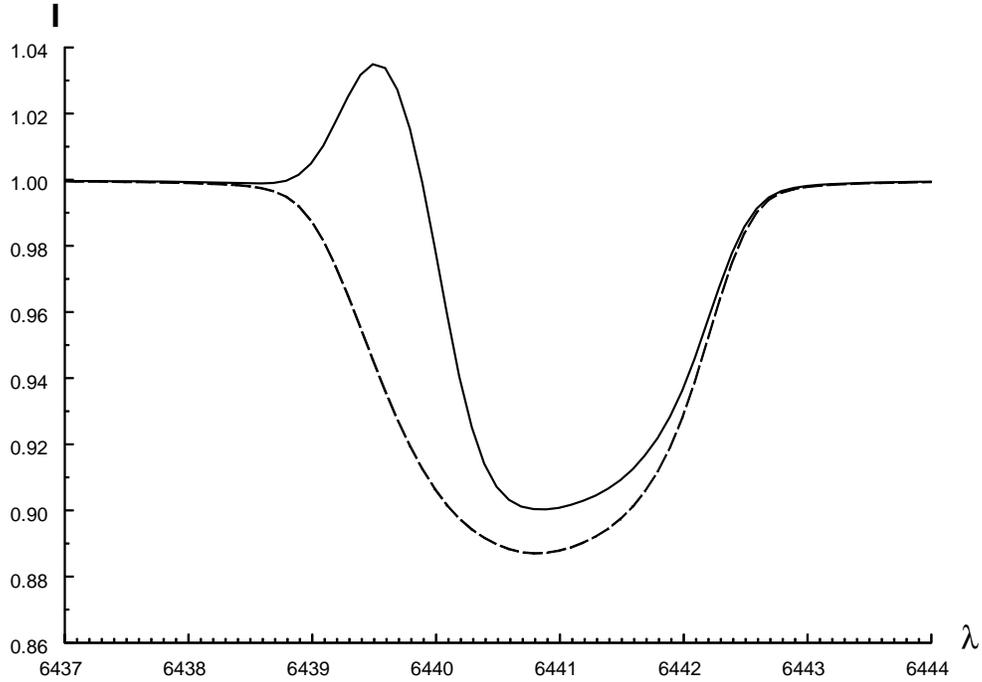}
\caption{Model integrated profiles of the CaI 6439\AA\, line of
the optical component in the 2S 0921-630 X-ray binary at orbital
phase 0.25. The profiles were obtained for the Roche model with
$m_x=2.35M_{\odot}$, $m_v=2.4M_{\odot}$, and $i=75^{\circ}$. The
dashed curve is the line profile derived in the Roche model
without including the reflection effect. The solid curve is the
line profile obtained with X-ray heating of the optical star
corresponding to $k_x=10$. Both model integrated profiles were
convolved with the spectrograph's instrumental profile, assuming
its FWHM=0.5\AA.} \label{CaIphase025}
\end{figure*}

Taking into account large orbital inclination, $i=75^{\circ}-
90^{\circ}$, and our relations between the masses of the
components in the 2S 0921--630 system (Fig. \ref{mxfmv}), we
conclude that the mass of the compact object is
$m_x\simeq1.55-2.55M_{\odot}$ if the optical star's mass is
$m_v=1.0-2.9M_{\odot}$. Our study demonstrates that, if we allow
for the possible mass loss by the optical component that reduces
its mass to $m_v=1M_{\odot}$, the presence of an accreting neutron
star with $m_x\simeq1.6-1.7M_{\odot}$ is most likely for this
binary. The observed X-ray spectrum \cite{Kallman2003} of the
compact object in 2S 0921--630 could correspond to a black hole,
as well as to a neutron star with a weak magnetic field. It
remains difficult to make an unambiguous selection between these
possibilities for the nature of the compact object.

Our study demonstrates that high-resolution spectrograms
($\lambda/\Delta\lambda=50000-100000$) are needed to estimate the
compact object's mass. The most correct approach to estimating
this mass, and hence revealing the nature of the compact object,
is to fit the orbital variations of absorption-line profiles in
the spectrum of the optical component.

We thank A.V. Tutukov for helpful comments.

\renewcommand{\bibname}{References}

\begin{figure*}
\vspace{0cm} \epsfxsize=0.99\textwidth
\epsfbox{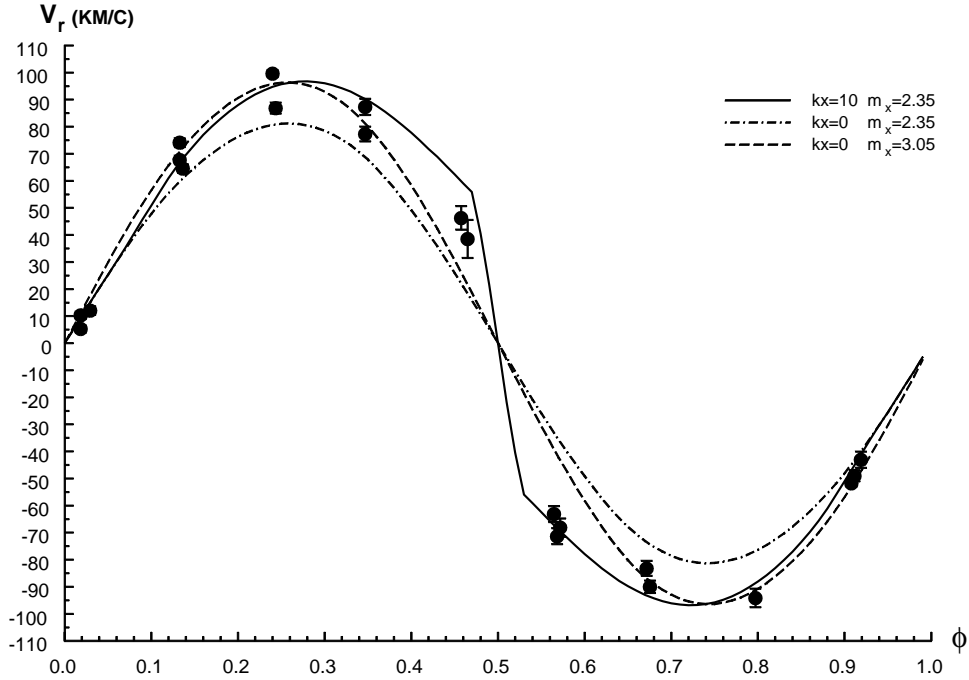} \caption{Observed
radial-velocity curve from \cite{Jonker2005} and the theoretical
radial-velocity curves obtained for the Roche model with i = 75.
Shown are the best-fit radial-velocity curves for X-ray heating
with $k_x=10$, with $m_v=2.4M_{\odot}$ and $m_x=2.35M_{\odot}$
(solid); without X-ray heating ($k_x=0$), with $m_v=2.4M_{\odot}$
and $m_x=2.35M_{\odot}$ (dash-dotted); and without X-ray heating
($k_x=0$), with $m_v=2.4M_{\odot}$ and $m_x=3.05M_{\odot}$
(dashed). The other parameters of the binary are given in Table
\ref{param Roche}.} \label{Vteor}
\end{figure*}

\begin{figure*}
\vspace{0cm} \epsfxsize=0.99\textwidth
\epsfbox{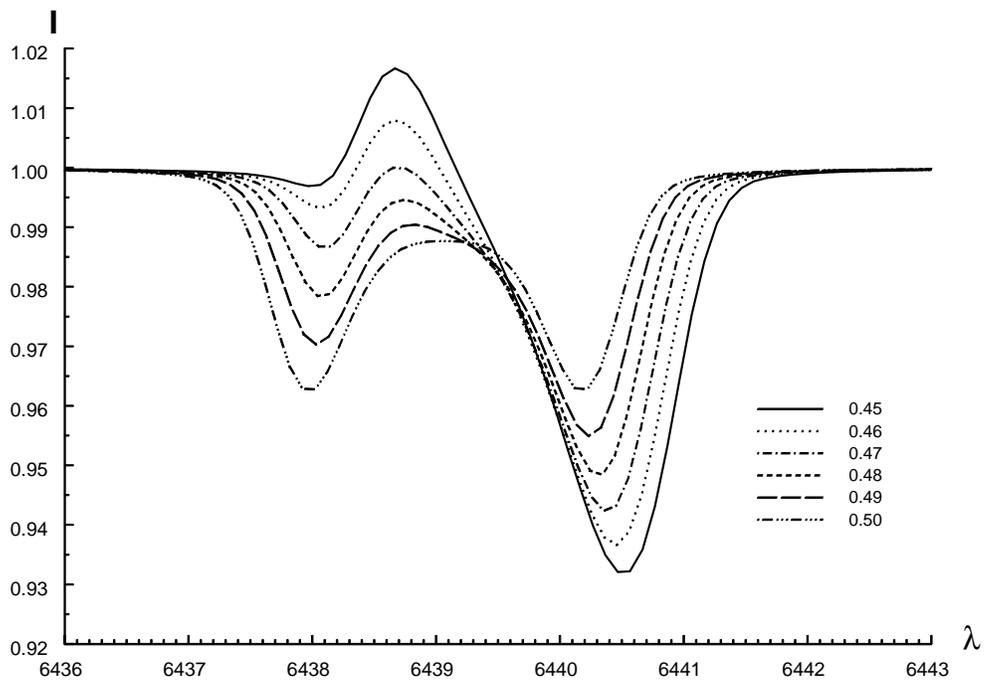} \caption{Model integrated
profiles of the CaI 6439\AA\, line obtained for the Roche model
with X-ray heating of the optical component ($k_x=10$) and for
$m_v=2.4M_{\odot}$, $m_x=2.35M_{\odot}$, $i=75^{\circ}$. The other
binary parameters are given in Table 1. Our notation for the
orbital phases is shown near the curves.} \label{CaIphase045050}
\end{figure*}

\begin{figure*}
\vspace{0cm} \epsfxsize=0.99\textwidth
\epsfbox{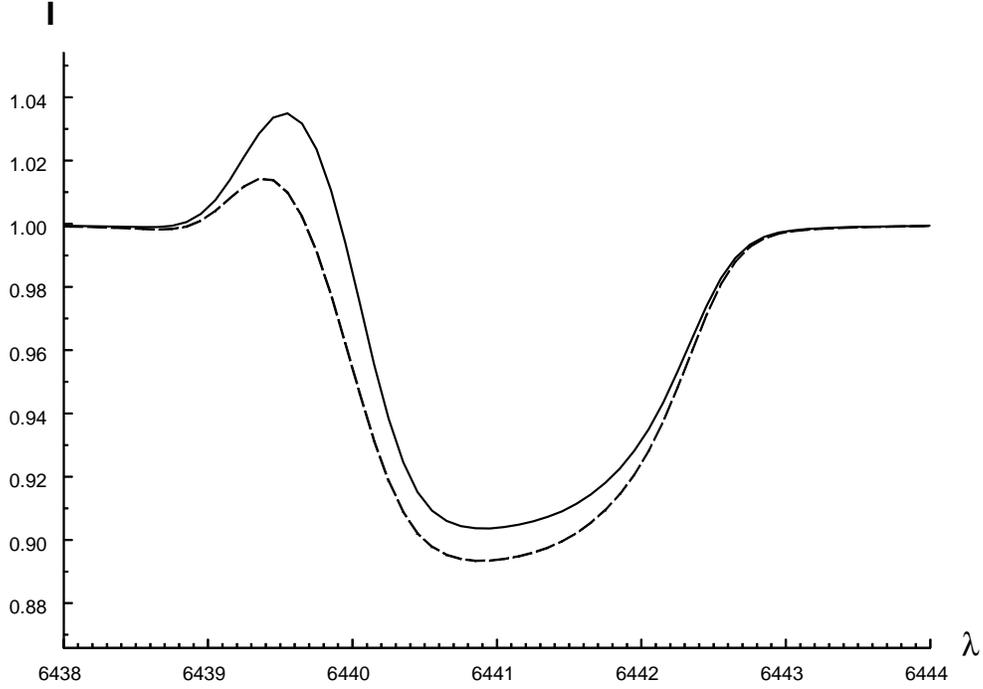}
\caption{Model integrated profiles of the CaI 6439\AA\, line of
the optical component in the 2S 0921-630 X-ray binary system at
orbital phase 0.25. The profiles were obtained for the Roche model
with $m_x=2.35M_{\odot}$, $m_v=2.4M_{\odot}$, $i=90^{\circ}$, and
$k_x=10$. Shown are the absorption-line profile assuming isotropic
(solid) and anisotropic (dashed) X-ray radiation from the compact
object. Both model integrated profiles were convolved with the
spectrograph's instrumental profile, assuming its FWHM=0.5\AA.}
\label{CaIphase025_ani}
\end{figure*}

\begin{figure*}[h!]
\vspace{0cm} \epsfxsize=0.99\textwidth
\epsfbox{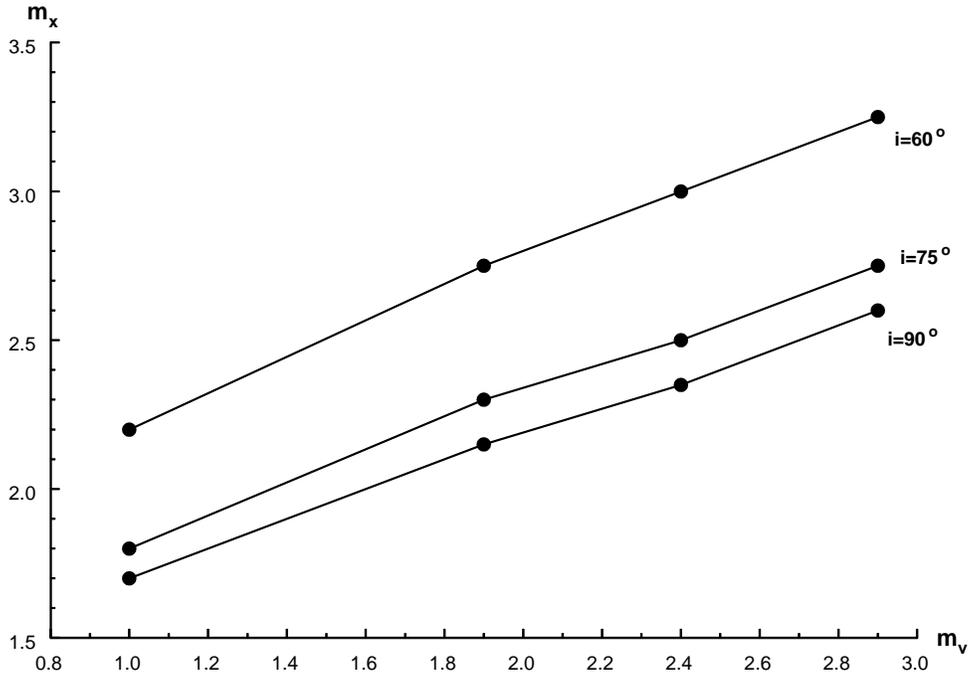} \caption{\small Relation
between $m_v$ and $m_x$ in the 2S 0921-630 binary obtained by
fitting the observed radial-velocity curve for the Roche model
with $k_x=10$ and anisotropic radiation from the accretion disk.
See text for details.}\label{mxfmv_ani}
\end{figure*}

\end{document}